\documentclass[3p,twocolumn,aps]{elsarticle}

\usepackage{epstopdf}

\usepackage{natbib}
\usepackage{dcolumn}

\usepackage{graphicx}
\usepackage{multirow}
\usepackage{subfigure}
\usepackage{tabularx}
\usepackage{varioref}
\usepackage[%
    colorlinks=true,%
    linkcolor=black,%
    citecolor=black,%
    urlcolor=black%
    ]{hyperref}
\usepackage[all]{hypcap} 
\usepackage{color}

\usepackage{footnote}
\usepackage{threeparttable}

\usepackage{amssymb}

\usepackage{amsmath}

\newcommand{\ra}[1]{\renewcommand{\arraystretch}{#1}}

\begin{document}

\begin{frontmatter}



\title{A new way of comparing double beta decay experiments}


\author[Milano,INFNMilano]{M.~Biassoni}
\author[INFNMilano]{O.~Cremonesi}
\author[roma2,LNGS]{P.~Gorla}

\address[Milano]{Dip.\ di Fisica dell'Universit\`{a} di Milano-Bicocca I-20126 - Italy}
\address[INFNMilano]{INFN Sezione di Mi-Bicocca, Milano I-20126 -Italy}
\address[LNGS]{Laboratori Nazionali del Gran Sasso, I-67010, Assergi (L'Aquila) - Italy}
\address[roma2]{INFN Sezione di Roma tor Vergata, I-00133, Roma - Italy}

\begin{abstract}
Many experiments whose goal is the search for neutrino-less double beta decay are taking data or in a final construction stage. The need for a tool that allows for an objective comparison between the sensitivity of different experiments is mandatory in order to understand the potential of the next generation projects and focus on the best promising technologies.

\end{abstract}


\end{frontmatter}


\section{Introduction}

Neutrino oscillation experiments have shown that neutrinos are massive particles that mix through the PMNS matrix.
On the other hand, the recent results 
showing that all the three mixing angles are different from zero, have opened a new window on the search for CP violations on the leptonic sector. 
Neutrinos have therefore proved to to be a powerful tool to point out the limitations of the Standard Model, demonstrating that new Physics beyond it must exists. 
Two very important neutrino properties are still missing in this framework: their nature and the absolute scale of their masses. Neutrino-less double beta decay provides a very effective way to find an answer to both of these questions. 
Indeed, presently available techniques for direct measurements of the electron antineutrino mass can only probe the quasi-degenerate mass region ($\delta$m $\ll$ m), while the much more sensitive cosmological observations suffer from heavy model dependances. 
	
Neutrino-less Double Beta Decay ($\beta\beta(0\nu)$) is a rare nuclear process in which a parent nucleus (A,Z) decays to a member (A,Z+2) of the same isobaric multiplet with the emission of two electrons: $ ^A_ZX \to ^A_{Z+2}X + 2e^-$). It violates the lepton number by two units and provides a powerful way to test neutrino properties. Indeed it can exist only if neutrinos are Majorana particles and can put important constraints on the neutrino mass scale. 

When the decay is mediated by the exchange of a light virtual Majorana neutrino, the $\beta\beta(0\nu)$ rate can be expressed as
	\begin{equation}\label{eq:rate}
	[ T_{1/2}^{0\nu}]^{-1} = 
	G^{0\nu}|M^{0\nu}|^2{\vert\langle m_{\nu} \rangle\vert^2}/m_e^2
	\end{equation}
where $G^{0\nu}$ is the phase space integral, $M^{0\nu}$ is the nuclear matrix element and $\langle m_{\nu} \rangle \equiv \sum_{k=1}^{3}\vert	U_{ek}^L \vert ^2m_ke^{i\phi _k }$ is a (coherent) linear combination of the neutrino masses weighted upon the mixing matrix. By the way, $\beta\beta(0\nu)$ provides also unique information on neutrino Majorana phases.

Altogether, the observation of $\beta\beta(0\nu)$ and the accurate determination of the $\langle m_{\nu} \rangle$ would  establish definitely that neutrinos are Majorana particles, fixing their mass scale and  providing a crucial contribution to the determination of the absolute neutrino mass scale.
However, even in the case that forthcoming $\beta\beta(0\nu)$ experiments would not observe any decay, important constraints could be obtained.  Indeed, assuming that neutrinos are Majorana particles, a negative result in the 20-30 meV range for $\langle m_{\nu} \rangle$ would  rule out the inverse ordering thus fixing the neutrino hierarchy problem. 
On the other hand, if future oscillation experiments would demonstrate the inverted ordering of the neutrino masses, a failure in observing $\beta\beta(0\nu)$ at a sensitivity of 20-30 meV would show that neutrinos are Dirac particles.
	
As can be easily deduced from Equation~(\ref{eq:rate}), the derivation of the only neutrino relevant parameter $\vert\langle m_{\nu} \rangle\vert$ from the experimental $\beta\beta(0\nu)$ results requires a precise knowledge of the transition Nuclear Matrix Elements M$^{0\nu}$(NME) for which many (unfortunately conflicting) evaluations are available in the literature\cite{faess11}. In fact, the spread in the available NME calculations causes a lot of confusion in the comparison of the results and of the expected sensitivities of the different experiments. 
Different approaches have been proposed for facing such a difficulty: i) reference to a single (arbitrarily chosen) calculation; ii) construction of a ``Physics Motivated Average'' (PMA) list of NME values~\cite{gome10}; iii) separate reference to all available calculations~\cite{crem10}.
In order to preserve correlations between different nuclei and allow a clearer comparison between the sensitivities of $\beta\beta(0\nu)$ experiments, we will refer here (when needed) to a single calculation~\cite{ibm2}, chosen just because it has the advantage of being available for all the nuclei of interest.

It's also worth noting that in a recent work~\cite{robertson}, hints of a possible anti-correlation between nuclear matrix elements and the phase space integrals of the same nucleus, have been suggested. Should this feature be confirmed, the relation between double beta decay rates and  neutrino Majorana masses (Equation~(\ref{eq:rate}) would become much simpler, allowing a more straightforward comparison between experiment using different $\beta\beta$ emitters, simply based on their half-lifetime sensitivities.
	
\section{Sensitivity}

Most sensitive experiments on $\beta\beta(0\nu)$ are based on counter methods for the direct observation of the two electrons emitted in the decay. They aim at collecting the limited available information (sum of the electron energies, single electron energy and angular distributions, identification and/or counting of the daughter nucleus) and are usually classified in {\em inhomogeneous} (when the observed electrons originate in an external sample) and {\em homogeneous} experiments (when the source of the $\beta\beta$ decay serves also as detector).
Both approaches are characterized by attractive features even if homogeneous experiments have provided so far the best results and characterize most of the future proposed projects.

In order to evaluate the potential of each experiment and compare it with the others, it is usual to refer to a detector {\em factor of merit} (or sensitivity), defined as the process half-life corresponding to the maximum signal n$_B$ that could be hidden by the background fluctuations at a given statistical C.L. 
At 1$\sigma$ level (n$_B$=$\sqrt{BTM\Delta}$), this is given by:
	\begin{eqnarray}
	\label{eq:sensitivity}
	F_{0\nu} = \tau^{Back.Fluct.}_{1/2}= \ln 2~N_{\beta\beta}\epsilon\frac{T}{n_B} = \nonumber \\
	= \ln 2\times \frac{x ~\eta ~ \epsilon ~ N_A}{A} \sqrt{ \frac{ M ~ T }{B ~ \Delta} } ~ (68\% CL)
	\end{eqnarray}  
where $B$ is the background level per unit energy and mass of detector, $M$ is the detector mass, $T$ is the measure time, $\Delta$ is the FWHM energy resolution, $N_{\beta\beta}$ is the number of $\beta\beta$ decaying nuclei under observation, $\eta$ is their isotopic abundance, $N_A$ the Avogadro number, $A$ the compound molecular mass, $x$ the number of $\beta\beta$ atoms per molecule, and $\epsilon$ the detection efficiency.

More sophisticated factors of merit have also been proposed~\cite{sens}. However, despite its simplicity (and high degree of approximation) Equation~(\ref{eq:sensitivity}) has the advantage of outlining all the relevant experimental parameters.

Among the parameters appearing in Equation~(\ref{eq:sensitivity}) the efficiency is probably the most delicate. 
Indeed it can depend strongly on the experiment details and the analysis method and some clarification is worth in order to avoid double counting. 
A fiducial volume in a TPC, for example, can be considered an efficiency that affects both the signal and the background. 
In cases like this it can be neglected in the evaluation of the sensitivity as far as the background is evaluated as the number of spurious events in the fiducial volume divided by the mass of the fiducial volume only. 
On the contrary, the escape probability of a double beta decay electron from the crystal surface in a bolometer is also an efficiency, but it reduces the number of signal events without affecting the background (which is still measured on the whole detector mass). The same holds for the signal efficiency of a shape cut in those detector where, during the data analysis, signal and background events can be discriminated based on the information associated to the event. This kind of efficiency de-facto reduces only the number of available double beta decay emitters for the detection and must be included in the $\epsilon$ term.

The way the parameter $B$ scales with the detector mass should also be briefly discussed. It is usually assumed to scale with the mass of the detector and this is usually a reasonable approximation for most of the experiments and detector technologies. 
In modular detectors, for example, the background is usually dominated by sources whose intensity is proportional to the surface of the detector. 
Since the detector is modular, however, the mass and the surface are also proportional. Indeed, in order to double the mass one has to double the number of sub-detectors, hence doubling the surface as well. 

On the contrary, in a noble gas detector (TPC, scintillator) the background is usually dominated by long range radioactivity (gammas and neutrons) because the surface radioactivity is easily rejected by imposing a fiducial volume in the analysis. In this case (at least for next generation detectors, where the volume is still not large enough to make the self shielding completely effective) the background directly scales with the volume (and mass) of the detector.

When the background level $B$ is so low that the expected number of background events in the region of interest (ROI), along the experiment life, is of order of unity ($MT\cdot B \Delta \sim$ O(1)), one generally speaks of a ``zero background''  (ZB)  experiment. Such a situation should common to a number of upcoming projects.
In these conditions, Equation~(\ref{eq:sensitivity}) can't be used anymore and a good approximation to the sensitivity is given by
	\begin{eqnarray}
	\label{eq:0sensitivity}
	F_{0\nu}^{\mathrm{ZB}} = 
	\ln 2~N_{\beta\beta}\epsilon\frac{T}{n_{\mathrm{CL}}} = \nonumber \\
	= \ln 2\times \frac{x ~\eta ~ \epsilon ~ N_A}{A} 
	\frac{ M ~ T }{n_{\mathrm{CL}}}
	\end{eqnarray}  

where $n_{\mathrm{CL}}$ is a constant depending on the chosen confidence level (CL) and on the actual number of observed events
. 

The most relevant feature of Equation (\ref{eq:0sensitivity}) is that F$_{0\nu}^{ZB}$ does not depend on the background level or the energy resolution and scales linearly with the active mass $M$ and the measure time $T$, therefore the sensitivity increases with the detector mass faster compared with the non-zero or finite background (FB) case.

The existence of two different regimes (ZB and FB) where the sensitivity shows a different dependence on the experimental parameters has striking consequences on the comparison between different experimental approaches. 
The two regions deserve a separate discussion and special attention has to be devoted to the crossing of their border.

In particular, it should be noticed that, for fixed values of $M$ and $T$, an improvement of the background $B$ or of the resolution $\Delta$ positively affects the sensitivity only as far as $B\Delta \gtrsim  1/MT$ (i.e. only in the FB region). 
In other words, the reduction of the background or the improvemet of the resolution at extreme values are useless if the mass is so small that the expected number of signal ($\beta\beta$) events is smaller than one (i.e. unchanged). 
In this case, only increasing the detector mass leads to an improvement of the sensitivity. 
Therefore the available statistics is the limiting factor and a well designed experiment should aim to match the condition $B \cdot \Delta \cdot M \cdot T \simeq 1$.

For most of the next generation high resolution calorimeters this corresponds to a limiting condition $B_{lim} \simeq \frac {1}{10\cdot M}$ or $B_{lim} \simeq 10^{-4}$ for a O(1t) experiment.

\section{Parameters redefinition}

Equations~(\ref{eq:sensitivity}) and (\ref{eq:0sensitivity}) define a quantity (the sensitivity) that should allow a simple direct comparison between different experiments. 
However, many parameters, like the isotopic abundance, the molecular mass and the number of atoms per molecule, appear in the definitions together with the more significant (and scalable) mass, resolution and background parameters. 
In order to compare in an exhaustive way experiments that use different techniques and materials, one should therefore analyse the sensitivity in a multi-dimensional space, taken into account all the relevant parameters. 
In the following we show a simple redefinition of the parameters that allows to reduce to two dimensions only the parameters space without loosing generality in the study.
First of all a parameter $\zeta$, characteristic of the experiment, can be defined as
\begin{equation}
\label{eq:zeta}
\zeta = \frac{x\eta\epsilon}{A}~.
\end{equation}
It gathers those parameters that are usually intrinsic properties of the experimental technique and the material used as source, and are therefore usually maintained when going from one detector generation to the next. 
Moreover,  as long as the efficiency $\epsilon$ is defined following the criteria previously discussed, the values of $\zeta$ and $M$ or $B$ are completely uncorrelated. The parameter $\zeta$ has the following dimensions:
\begin{eqnarray}
[\zeta] & = & \frac{\mathrm{\#~of~moles~of~``efficient''~} \beta\beta \mathrm{~isotope}}{\mathrm{mass}} \nonumber \\
& = & \frac{\mathrm{n}_{\beta\beta}}{\mathrm{kg}} \nonumber
\end{eqnarray}
Equation~(\ref{eq:sensitivity}) then becomes:
\begin{eqnarray}
\label{eq:sensitivity corr}
F_{0\nu} & = & \ln 2 ~ N_{A} \times \zeta \sqrt{ \frac{ M ~ T }{B ~ \Delta} } = \nonumber \\
& = & \ln 2 ~ N_{A} \times \sqrt{ \frac{ \zeta M ~ T }{\frac{B}{\zeta} ~ \Delta} } = \nonumber \\
& = & \ln 2 ~ N_{A} \times \sqrt{ \frac{ \tilde M ~ T }{\tilde B ~ \Delta} }
\end{eqnarray}

and analogously Equation~(\ref{eq:0sensitivity}) can be written as:

\begin{eqnarray}
\label{eq:0sensitivity corr}
F_{0\nu} & = & \ln 2 ~  N_{A} \times \frac{\zeta M T}{n_{L}} = \nonumber \\
& = & \ln 2 ~  N_{A} \times \frac{\tilde M T}{n_{L}} ~.
\end{eqnarray}

In both Equation~(\ref{eq:sensitivity}) and Equation~(\ref{eq:0sensitivity}) the substitutions
\begin{equation}
\zeta M = \tilde M  \:\:\:\:\:\:\:\:\:\:\: \mathrm{and}  \:\:\:\:\:\:\:\:\:\:\:  \frac{B}{\zeta} = \tilde B
\end{equation}

have been performed. 
Given the dimensions of $\eta$, the new $\tilde M$ parameter represents the number of moles of $\beta\beta$ emitting isotope in the detector, while $\tilde B$ is the background expressed in counts per unit energy per mole of emitting isotope per year. 
The sensitivity \emph{factor of merit} keeps, obviously, the same meaning and dimensions, but, by redefining both the detector mass and its background in terms of the number of moles of emitting isotope, it can be expressed in terms of a smaller subset of (explicit) parameters. It should be noticed that by redefining the detector mass and background, the separation between the two sensitivity regimes (ZB and FB) is also conserved, since the parameter re-definition does not affect the condition
\begin{equation}
B \Delta \cdot M T = \tilde B \Delta \cdot \tilde M T
\end{equation}

 The situation can be further simplified by comparing different experiments for the same measure time $T$ (which is a reasonable assumption because the run time of $\beta\beta$ experiments is usually fixed by external constraint to some period not very different from 5 years):
\begin{equation}
\label{eq:PS}
S = \tilde M\cdot T  \:\:\:\:\:\:\:\:\:\:\: \mathrm{and}  \:\:\:\:\:\:\:\:\:\:\:  P = \tilde B \cdot \Delta ~.
\end{equation}
$S$ will be called the \emph{scale} of the experiment (expressed in units of n$_{\beta\beta}\cdot$y), while $P$ is the expected number of background counts in the ROI per mole of $\beta\beta$ isotope per year, that will be called \emph{performance} and expressed in units of cnts/n$_{\beta\beta}$/y hereafter.

With these two final variables Equations~(\ref{eq:sensitivity corr}) and (\ref{eq:0sensitivity corr}) becomes
\begin{equation}
\label{eq:sensitivity corr2}
F_{0\nu}  =  \ln 2 ~  N_{A} \times \sqrt{ \frac{S}{P} }
\end{equation}
and
\begin{equation}
\label{eq:0sensitivity corr2}
F_{0\nu}^{\mathrm{ZB}}  =  \ln 2 ~  \frac{N_{A}}{n_{L}} \times S
\end{equation}
or, remembering the condition that determines the separation between the two regimes,
\begin{equation}
\label{eq:final}
F_{0\nu} = \begin{cases} 
\ln 2 ~  N_{A} \times \sqrt{ \frac{S}{P} }, & \mbox{if } P\cdot S >1 \\ 
\ln 2 ~  \frac{N_{A}}{n_{L}} \times S, & \mbox{if } P \cdot S \lesssim1 
\end{cases}
\end{equation}

Equation~\ref{eq:final} is a simple but exhaustive form of the sensitivity that allows:
\begin{itemize} 
\item to quickly and realistically compare two experiments once the specific values of $P$ and $S$ are calculated;
\item to understand which detector features are worth improving in order to effectively increase the sensitivity.
\end{itemize}

\section{Critical comparison}

In the $(P,S,F_{0\nu})$ space each experiment sensitivity represents a point laying on the surface described by Equation~\ref{eq:final}. 
In Figure~\ref{fig:movements lin} the projection of the surface on the $(P,S)$ plane is shown; the black lines are iso-sensitivity curves, while the yellow line highlights the boundary region between the two regimes: in the left region the sensitivity doesn't depend on the $P$ parameter (background and resolution) anymore, but steeply (linearly) varies increasing the \emph{scale} $S$.

\begin{figure}[h!]
	\label{fig:2Dlog}
	\includegraphics[width=0.95\columnwidth]{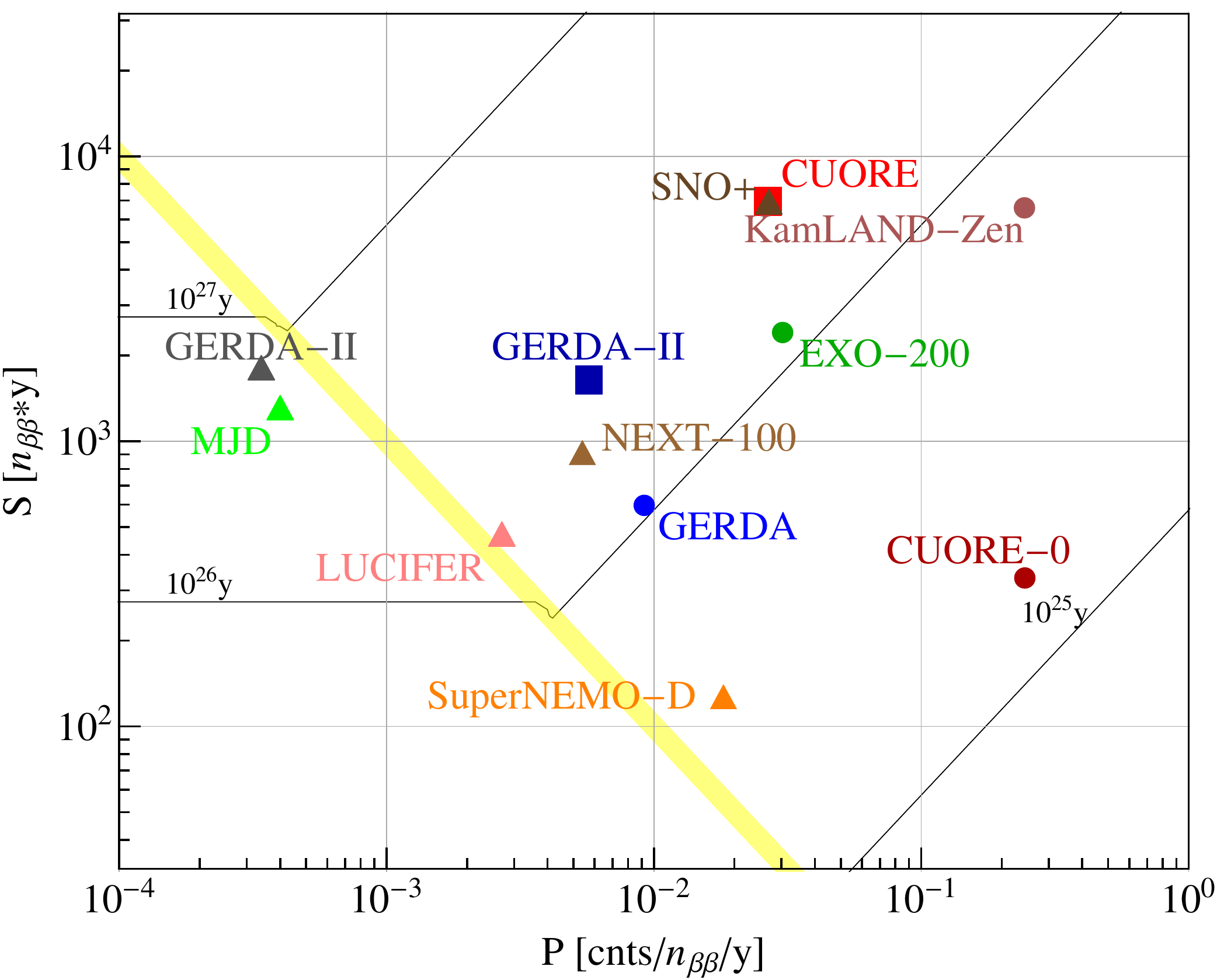}
	\caption{$P-S$ plane with iso-sensitivity curves, log scale. Some of the most important present and future $\beta\beta$ experiments are reported: point = running experiment; square = realistic estimation; triangle = conceptual design.}
\end{figure}

\begin{figure*}[t]
	\label{fig:3Dlog}
	\centering
	\includegraphics[width=0.9\textwidth]{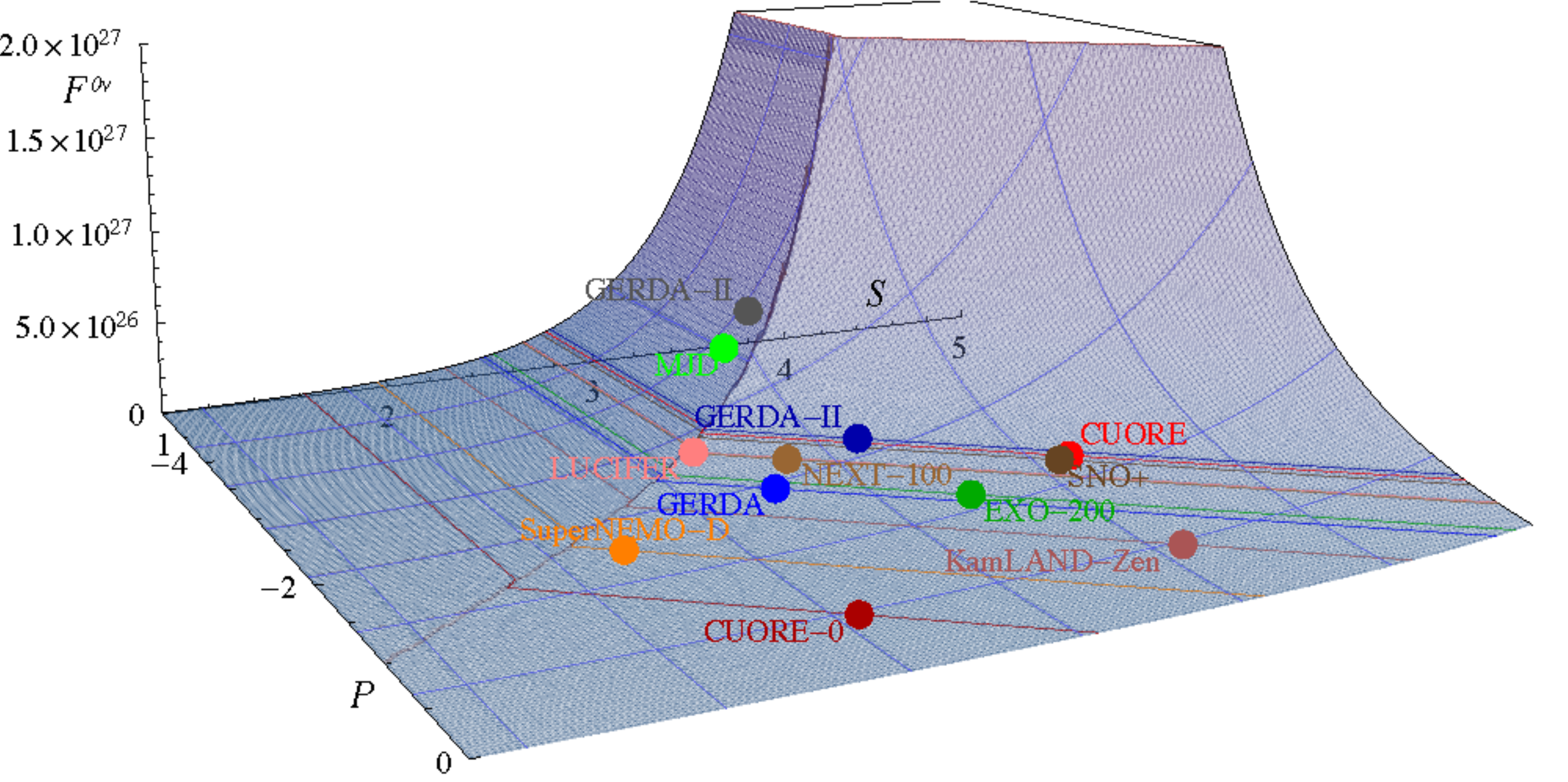}
	\caption{Same experiments of Figure~\ref{fig:2Dlog} on the sensitivity surface.}
\end{figure*}

Using logarithmic scales for $P$ and $S$ axes the boundary between the two regimes, which is simply defined, in this parameter space, by the condition $P\cdot S = 1$, is a line with negative slope -1. The iso-sensitivity curves are also lines, parallel to the $x$ axis in the ``zero background'' region and with unitary slope in the other region. In Figure~\ref{fig:2Dlog} some of the most important current and future $\beta\beta$ experiments are reported in the two-dimensional plot, while Figure~\ref{fig:3Dlog} represents the same data but with the experiments laying on the sensitivity surface. The numerical values of the parameters used for the comparison are summarized in Table~\ref{tab:exp}~\cite{crem13}.

\begin{table*}[tb]
	\centering
	\small
	\ra{1.2}
	\begin{threeparttable}
		\begin{tabular}{cccccccc}
			\hline\hline
			Experiment & Isotope & $\tilde M$ & $\tilde B [\times10^{-3}]$ & $\Delta$ & $P [\times10^{-3}]$ & $S$(5y) & $F_{0\nu} [\times 10^{26}$y] \\
			\hline

			CUORE\cite{CUORE1,CUORE2} & $^{130}$Te & 1389.5 & 5.3 & 5 & 26.7 & 6947.5 & 2.13\tnote{b} \\
			CUORE-0\cite{CUORE1} & $^{130}$Te & 66.3 & 44 & 5.6 & 244 & 331.5 & 0.15\tnote{a} \\
			GERDA\cite{GERDA} & $^{76}$Ge & 119.2 & 1.9 & 4.8 & 9.2 & 596 & 1.06\tnote{a} \\
			GERDA-II\cite{GERDA} & $^{76}$Ge & 328.2 & 1.8/0.11 & 3.2 & 5.7/0.34 & 1641 & 2.24\tnote{b} /6.01\tnote{c} \\
			KamLAND-Zen\cite{KAMLAND} & $^{136}$Xe & 1318.2 & 1.0 & 243.2 & 243.2 & 6591 & 0.69\tnote{a} \\
			EXO-200\cite{EXO} & $^{136}$Xe & 481.6 & 0.31 & 96.5 & 30.3 & 2408 & 1.18\tnote{a} \\
			MJD\cite{MJD} & $^{76}$Ge & 237.6 & 0.095 & 4 & 0.4 & 1188 & 4.35\tnote{c} \\
			SuperNEMO-D\cite{SUPERNEMO} & $^{82}$Se & 23 & 0.15 & 120 & 18.2 & 115 & 0.33\tnote{c} \\
			SNO+\cite{SNO} & $^{130}$Te & 1252.8 & 0.11 & 240 & 26.9 & 6264 & 2.01\tnote{c} \\
			NEXT-100\cite{NEXT} & $^{136}$Xe & 165.4 & 0.44 & 12.5 & 5.4 & 827 & 1.63\tnote{c} \\
			Lucifer\cite{LUCIFER} & $^{82}$Se & 125.1 & 0.2 & 20 & 4 & 636.5 & 1.65\tnote{c} \\

			\hline\hline
		\end{tabular}
		\begin{tablenotes}
			\item[a] \tiny{The experiment is running and the parameters have been measured}
			\item[b] \tiny{The experiment feasibility has been demonstrated and the parameters values have been measured with demonstrators (realistic estimation)}
			\item[c] \tiny{The experiment is in a conceptual design phase; the parameters values are theoretical estimations}
		\end{tablenotes}
	\end{threeparttable}
	\caption{Experimental parameters of the the most evolved experiments. The units of the background $\tilde B$ are cnts/keV/n$_{\beta\beta}$/y, those of the mass $\tilde M$ are n$_{\beta\beta}$ and the resolution $\Delta$ is in keV.  Scale $S$ is expressed in units of n$_{\beta\beta}\cdot$y and performance $P$ in units of cnts/n$_{\beta\beta}$/y. The sensitivity is for 5 years of data taking.}
	\label{tab:exp}
\end{table*}

\section{Movements on the $P-S$ plane}
\begin{figure}[t]
	\label{fig:movements}
	\includegraphics[width=0.9\columnwidth]{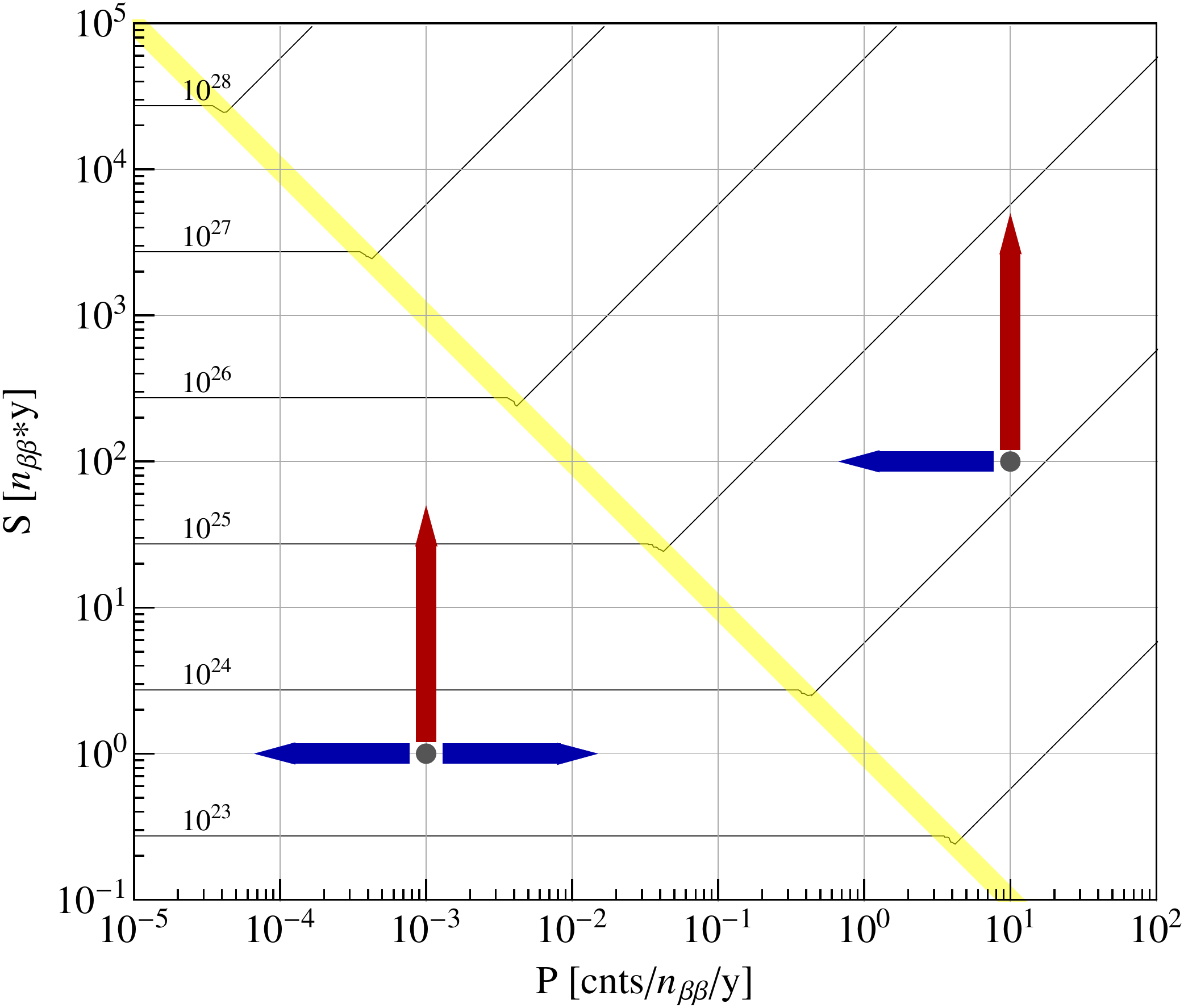}
	\caption{Movements of an experiment on the $P-S$ plane, when the background level is changed (blue arrows) or the detector mass is increased (red arrows).}
\end{figure}

In Figure~\ref{fig:2Dlog} different experiments are reported in order to directly compare their sensitivity to neutrino-less double beta decay. The same plot, however, can be used to understand which is the most effective strategy for an experiment (or, better, an experimental technique) in order to improve its sensitivity while minimizing the R\&D or economical effort. 
In Figure~\ref{fig:movements}, the arrows represent the directions an experiment would move along by changing some of the relevant experimental  parameters, embedded in the new definition of the variables. 

\begin{figure}[t]
	\label{fig:movements lin}
	\includegraphics[width=0.9\columnwidth]{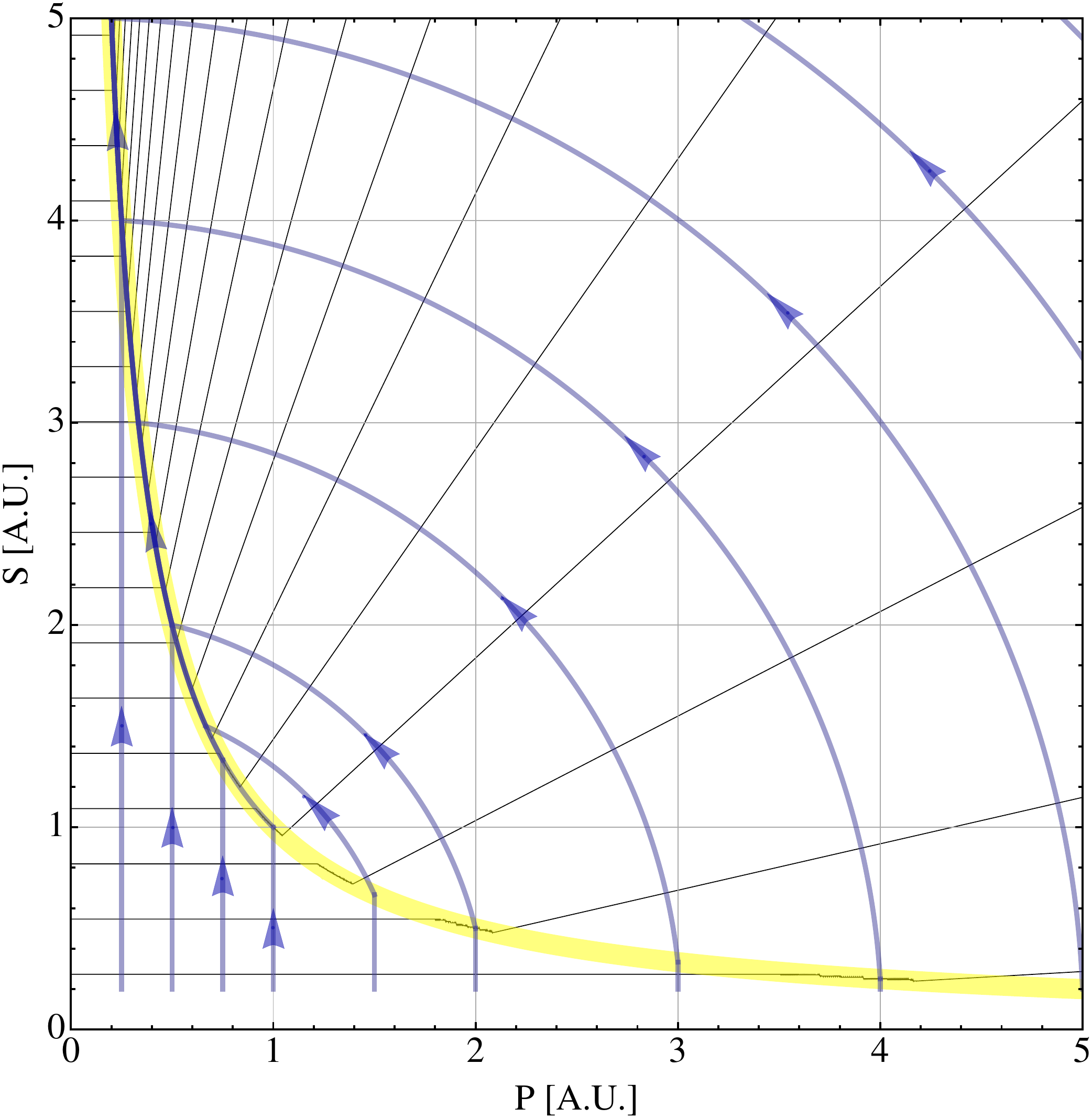}
	\caption{Paths in the $P-S$ plane corresponding to the steepest increase of the sensitivity. The paths (blue lines and arrows) are perpendicular to the iso-sensitivity curves (black lines): they are vertical lines in the zero background region and circles in the non-zero background region. All the curves intercept the boundary between the two regimes and from that point the boundary itself is the steepest path.}
\end{figure}

The blue arrows represent the effect of a variation in the background level ($B$). The experiment position shifts along the $P$ axis because the background enters in the definition of this variable only. In the top-right region a reduction of the background leads to an improvement of the sensitivity; a variation of the background in the bottom-left region of the plot, corresponding to the zero-background regime, doesn't affect the sensitivity, as expected, because the corresponding movement is parallel to the iso-sensitivity curves. 
Therefore, once an experiment enters the zero-background region, a further improvement of the background level alone is useless. On the contrary, an increase of the detector mass (red arrows) shifts the experiment along the $S$ axis, improving the sensitivity in both regions of the plot.
In general, however, the path that maximises the sensitivity improvements depends on the region of the plot where the experiment lies, and it is defined as the gradient of the sensitivity or, visually, as the perpendicular to the iso-sensitivity curves in each point of the plane. Figure~\ref{fig:movements lin} represents this paths as blue lines, and the arrows show the direction of increasing sensitivity. In the zero background region, where the sensitivity doesn't depend on the \emph{performance}, the optimal path is parallel to the \emph{scale} axes, as expected: only an increase of the number of emitting nuclei can improve the sensitivity. On the right of the boundary between the two regimes the iso-sensitivity curves are defined by Equation~\ref{eq:sensitivity corr2} that explicitly reads
\begin{equation}
S = K ~ P
\end{equation}
with $K=\left(\frac{F_{0\nu}}{\ln 2 N_{A}}\right)^{2}$; the iso-sensitivity curves are therefore lines with zero intercept and the slope depending on the sensitivity value, and the steepest paths are circles.
The implications are interesting: all the paths, whether they start from the zero or the non-zero background region, intercept at some point and afterwards move along the boundary between the two regimes, that we therefore define the \emph{golden region}. 

\begin{figure}[t]
	\label{fig:movements log}
	\includegraphics[width=0.9\columnwidth]{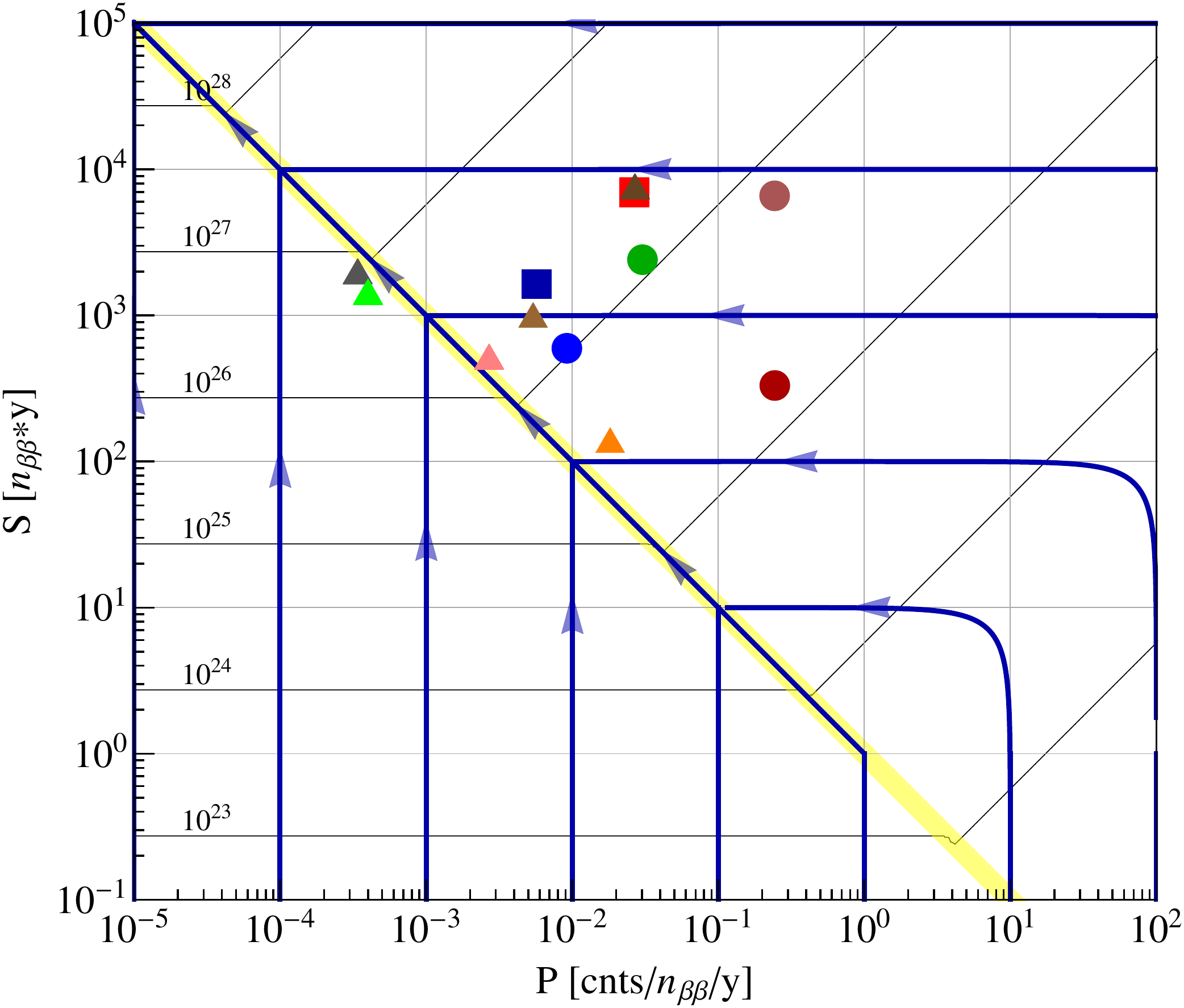}
	\caption{Same as Figure~\ref{fig:movements lin} in the usual log-log scale. The same experiments represented in Figure~\ref{fig:2Dlog}, with the same shape and colour code for the markers, are reported.}
\end{figure}

By representing the same plot in log-log scale (Figure~\ref{fig:movements log}) the steepest paths are not perpendicular to the iso-sensitivity curves anymore (the transformation doesn't conserve the angles), but the same features are still evident:
\begin{itemize} 
\item in the zero background region the parameter that is affecting the sensitivity is the \emph{scale} only;
\item in the non-zero background regime both \emph{performance} and \emph{scale} need to be improved. Depending on the position in the plane both one or the other can be the most important parameter; in general, the larger is the \emph{scale}, the more important becomes the \emph{performance} of the experiment, while for a poor \emph{performance} experiment the sensitivity dependance on the \emph{scale} is stronger; basically all the present and future experiments with non-zero background lie in a region of the space where the emph{performance} is the most critical parameter to be improved in order to reach the golden region;
\item once an experiment reaches the golden region, the largest sensitivity increase is obtained by improving by the same factor both the \emph{performance} and the \emph{scale} (improving the \emph{performance} means reducing its magnitude while the opposite holds for the \emph{scale}, of course); this effect is obtained, for instance, by increasing the isotopic abundance.
\end{itemize}

\section{Conclusions}

Together with the possibility of performing a credible comparison between the $\beta\beta 0 \nu$ sensitivities of different experiments, the approach shown in this paper gives a plain indication on the best approach the various experiments should go through in order to improve their sensitivity, or, in a figurative meaning, to climbing the slope of Figure~\ref{fig:3Dlog}. In particular, reducing the background and the energy resolution is worth the effort only as long as an experiment lays on the rightmost region of the plot. Once the boundary between the two regimes has been crossed, an improvement of the sensitivity can be obtained only by increasing the number of $\beta\beta$ emitter nuclei. 







%

\end{document}